\documentclass[showpacs,preprintnumbers,amsmath,amssymb,aps,twocolumn]{revtex4}

\usepackage{graphicx}
\usepackage{dcolumn}
\usepackage{bm}
\begin{document}

\title{High resolution portrait of the ideal trefoil knot}
\author{Sylwester Przybyl and Piotr Pieranski}
\affiliation{Laboratory of Computational Physics\\ 
Poznan University of Technology \\ 
Nieszawska 5B/11, 60-965 Poznan, Poland.} 
\email{Sylwester.Przybyl@put.poznan.pl}
\thanks{This work was carried out under Project 06/62/DSPB/0214/2014.}

\date{\today}

\begin{abstract}
The shape of the most tight trefoil knot with $N=200640$ vertices found with an appropriately modified finite element method is analyzed. The high number of vertices makes plots of its curvature and torsion very precise what allows the authors to formulate new, firmly justified conjectures concerning the shape of the ideal trefoil knot.

\end{abstract}
\maketitle
\section[Introduction]{Introduction}
Knots tied in a rope change their shape when tightened by making the rope shorter. Conformations at which the length/diameter ratio reaches its global minimum are often referred to as {\it ideal}. Although a lot of work has been devoted to the problem, the ideal conformations of knots even as simple as the trefoil knot are only vaguely  known today. It is the aim of the present paper to describe results of a numerical simulation that allowed us to generate a portrait of the ideal trefoil knot at such a high resolution that new, essential details of its shape become clearly visible. Some of the details look at the first sight unusual. One may suspect that they are but artifacts of the discrete nature of the simulated knot. To exclude this possibility we performed, at a similar precision, numerical simulations of the simple clusp - the system whose geometrical details have been discovered analytically. As we shall demonstrate our simulation method is able to reproduce the details. Thus, the similar subtle details that we see in the structure of the tight trefoil knot are by all means reality and not artifacts of our simulation method.

\section[DiscretizationProblems]{The continuous trefoil knot and problems with numerical simulations of its tightening}

The tightening of the closed knots needs applying a hypothetical process, in which the length of the rope diminishes while the diameter of its perpendicular sections remains constant. Let us note that for a closed knot its length is defined simply as the length of the rope in which it is tied. The longitudinally shrinking rope gets into contact with itself. When the surface of the rope is slippery, and when the rope is hard, the self-contact points move along its surface without any hindrance. As a result, the knot changes its shape until a conformation is reached at which the tightening process stops. Once more, computer experiments indicate that for larger closed knots one may expect many tight conformations to which one arrives starting from different initial conformations \cite{CantarellaCompositeKnots}.    

Physical properties of the perfect rope induce particular geometrical properties of knots that are tied in it. The assumption, that the rope is hard, i.e. its circular sections are not deformable and cannot overlap, implies that its axis has a continuous, but not necessarily smooth, tangent vector. As a result, its curvature versus the arclength function $\kappa(l)$, must remain free from any Dirac-delta-like components, but it needs not to be continuous. Obviously, as explained above, it cannot be larger than $1/R$. 

The hypothetical tightening process performed on the trefoil knot tied in the perfect rope has been simulated in the past using different numerical procedures. Unfortunately, such simulations are always awkward, since the unavoidable discretization of the knot results in dealing not with the knot itself but with its discrete, polygonal representation. 
Let $N$ be the number of vertices in the polygonal representation $K_p$ of the simulated knot $K$. At the final analysis of the most tight conformation obtained from the numerical simulation, the tight polygonal knot $K_p$ is replaced by a $C^1$ smooth knot $K_c$ built from circular arcs inscribed into edges of the polygon. It seems natural to expect that as $N$ becomes larger, the knot inscribed into the most tight polygonal knot approximates better the ideal conformation ${\cal K}_{id}$. The toll paid for the discretization is high: at its full resolution, the curvature of the knot inscribed into numerically found polygonal knot displays some features which are certainly not present in the ideal conformation and must be seen as artifacts of the discretization. To get rid of them and see the true shape of the knot, the resolution must be brought down. As a result,  the image of the knot becomes more vague, but this allows us to see better the true shape of the ideal conformation. When $N$ is very large (of order $10^5$), lowering resolution 10 times gives an image still precise enough to reveal subtle details of the knot shape. 

\section{Numerical simulation of the simple clasp} \label{Clasp}

\begin{figure}
	\centering
		\includegraphics[width=200pt]{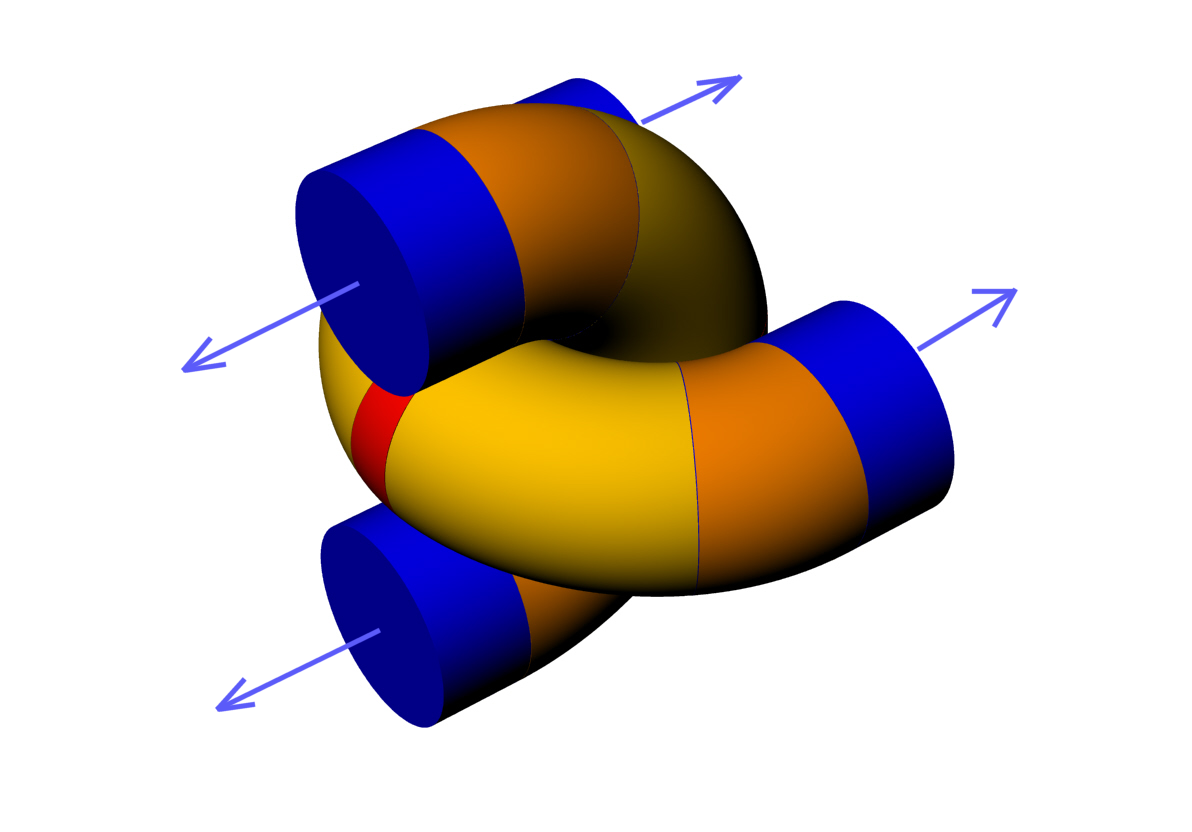}
		\includegraphics[width=200pt]{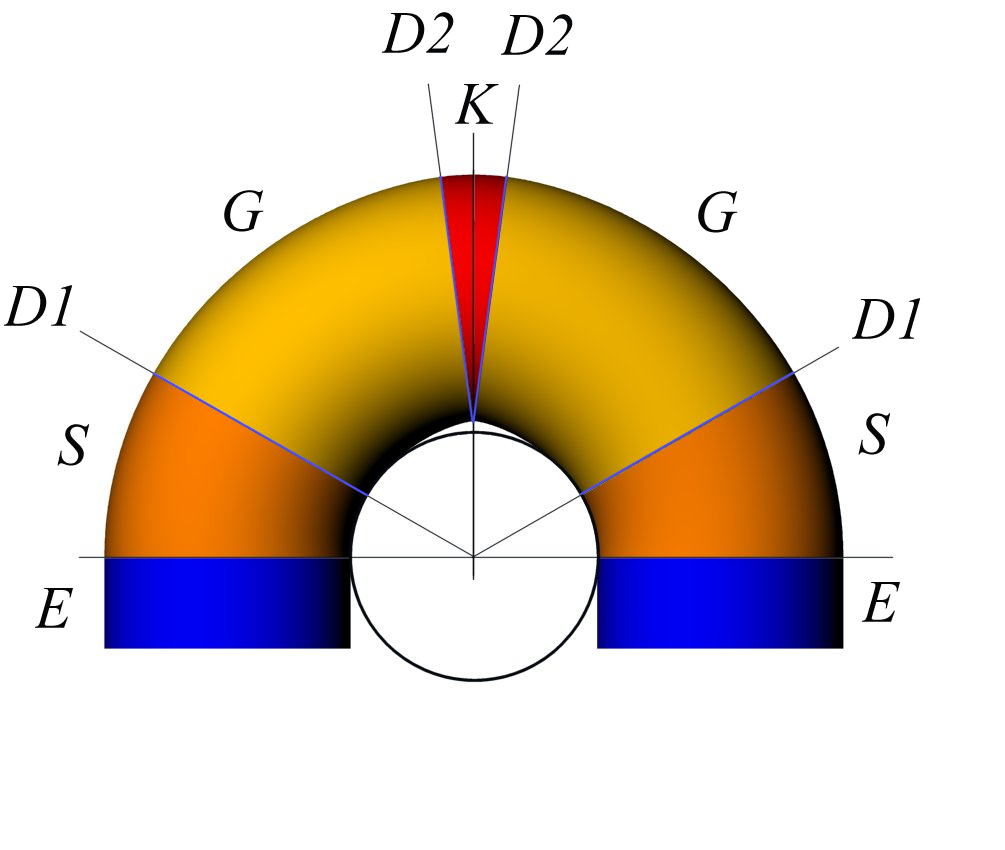}
	\caption{The most tight simple clasp found by the SP-FEM method and the details of its anatomy: {\it E}-straight ends, {\it S}-shoulders, {\it G}-Gehring, {\it K}-kink, {\it D1}-first deeps, {\it D2}-second deeps. See text. }
	\label{SimpleClasp}
\end{figure}

\begin{figure}
	\centering
		\includegraphics[width=250pt]{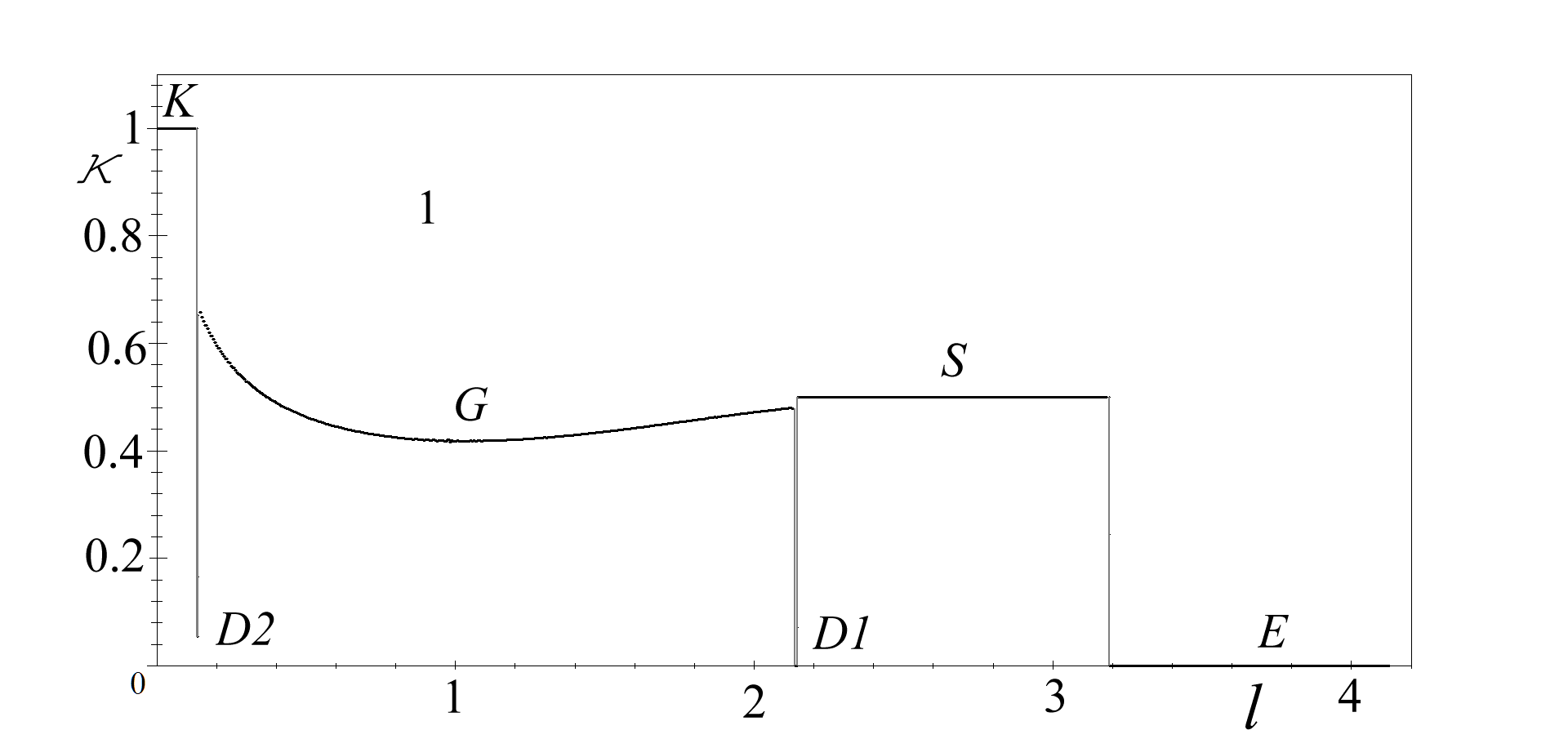}
		\caption{Curvature of the most tight simple clasp found by the SP-FEM simulation method. The plot stays in a quantitative agreement with the plot presented by Starostin \cite{Starostin2003}.}
	\label{SimpleClaspCurvature}
\end{figure}

\begin{figure}
	\centering
		\includegraphics[width=250pt]{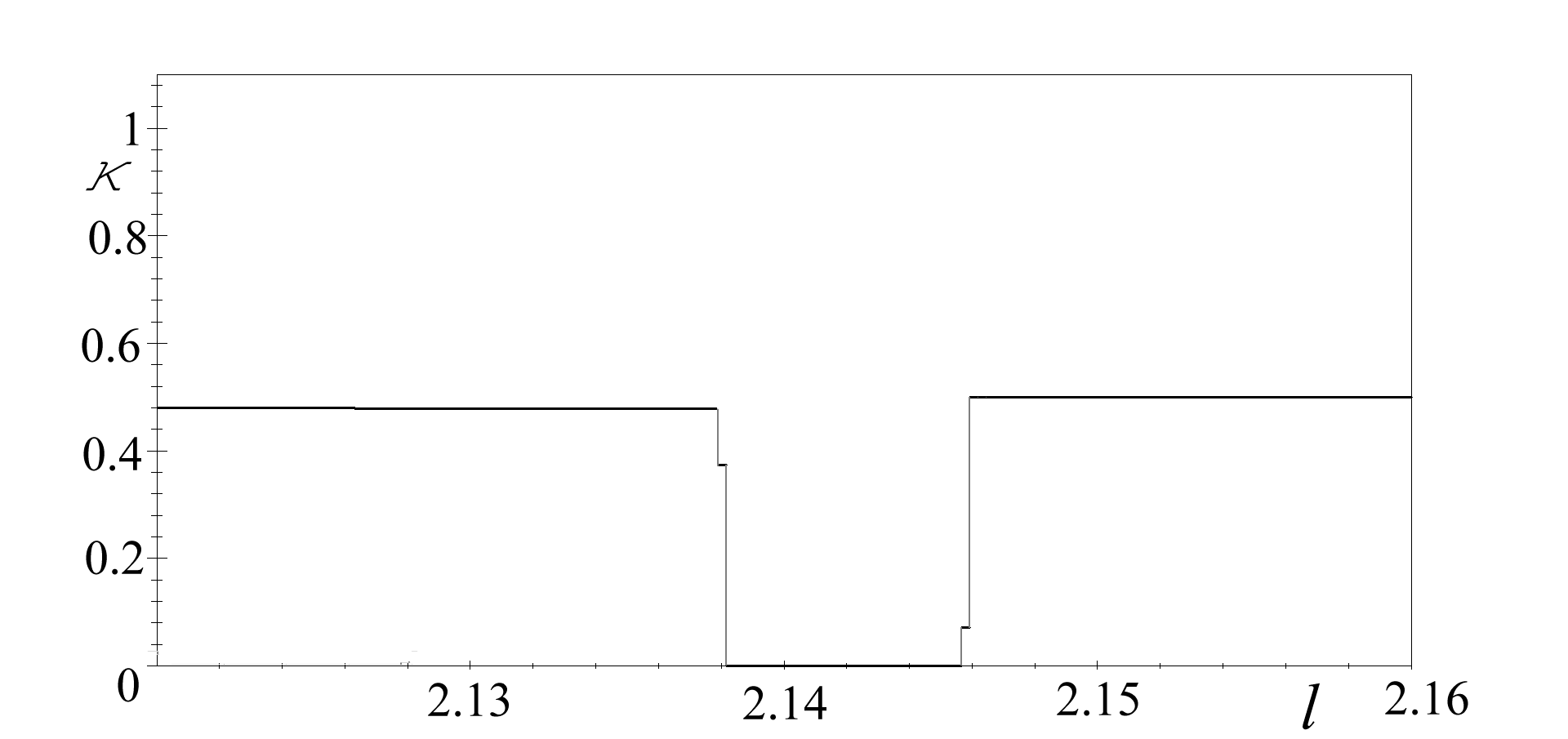}
		\caption{Curvature in the {\it D1} region of the simple clasp. The length of the straight piece equals 0.008, what stays in a quantitative agreement with Starostin's solution \cite{Starostin2003}.}
	\label{SimpleClaspCurvatureD1}
\end{figure}

As said above, performing numerical experiments we are constrained to work only with finite resolution portraits of the ideal conformation. Analyzing the portraits we are trying to guess, what the shape of the ideal conformation is. A much better choice would be finding the ideal conformation via a rigorous analytical calculation. Unfortunately, nobody knows so far how to perform such a calculation even for a knot as simple as the trefoil knot. (The ideal conformation is rigorously known only for the closed unknot, but here no calculation is needed: the ideal unknot is a circle of radius $R$.) As indicated to us by Starostin and Sullivan, the analytical analysis path proved to be successful in the case of the so-called {\it simple clasp}.

Connect two perpendicular ropes and pull away their ends as shown in Fig. \ref{SimpleClasp} a until the clasp becomes tight. In spite of what our intuition suggests, the structure of the simple clasp is by no means simple. This was demonstrated by Starostin and Sullivan \cite{Starostin2003}, \cite{Sullivan2001}. In what follows we shall refer to the structure as the {\it SS solution}. Starostin \cite{Starostin2003}, using a variational approach, demonstrated that to minimize its length, each of the ropes involved in the clasp must take a particular shape consisting of a few curved pieces separated from each other by extremely short straight pieces. Cantarella et al. solved the problem in a more general context demonstrating that the multi-piece solution is critical \cite{Cantarella2006}, \cite{Cantarella2011}. Figure \ref{SimpleClasp}b presents the details of the solution. Let us describe a representative part of it.

The easy part of the solution is the shape of the end parts of the rope, i.e. those pieces of it that are not in touch with the other rope. Being free, they must be straight. The pieces are marked by {\it E}. The next piece is what SS call {\it shoulder} ($S$). Here, one rope stays in touch with the other rope along an arc of its central circular section. Thus, curvature of the shoulder piece is equal $1/2$. (We assume that the rope radius $R=1$.) The shape of the next curved piece is not so obvious. SS call it {\it Gehring} ($G$), since this part of the clasp can be seen as a fragment of the general solution of the so-called Gehring problem, in which there are no limits on the curvature of the clasped ropes. Within the Gehring piece the rope stays in a double touch with the other rope along two curved lines. Curvature changes here in a smooth manner. The last curved piece of the simple clasp structure is located at the center. SS call it {\it kink} ($K$). Within the kink piece, the rope stays in touch with itself. This happens at a single point, where circular sections of the rope are tangent to each other. Curvature reaches here its highest allowable value $\kappa=1$. Parts $E$, $S$, $G$ and $K$ are of comparable sizes and are easy to find in a numerical simulation. The problem is that, as SS revealed, both between $S$ and $G$, as well as between $G$ and $K$, there are two additional, extremely short pieces $D1$ and $D2$, where the rope becomes straight, i.e. where curvature drops to zero. The length of the pieces is so small that in figures where SS presented their solutions, they could not be visualized in a proper manner. Pieces $D1$ and $D2$ of the SS solution are extremely short, but it does not mean we can neglect them. On the contrary, they should be seen as the fingerprints of the ideal conformation of the simple clasp. If they are not seen in a numerical simulation of the clasp, than either the simulated clasp is not tight, or the simulation is not precise enough.
The SP-FEM simulation was subject to this precision test. In the simulation each of the clasp ropes was represented by a polygonal curve containing $N=31992$ vertices. The plot of the curvature of a half of one of the ropes is presented in Fig. \ref{SimpleClaspCurvature}. As one can see, all pieces predicted by the SS solution are there, although the $D2$ piece is not properly developed. (Our figure \ref{SimpleClaspCurvature} can be confronted with Fig. 20 in ref. \cite{Starostin2003}.) The details of the curvature plot with the region of the $D1$ piece are presented in Fig. \ref{SimpleClaspCurvatureD1}. As seen in the figure, the width of the $D1$ piece equals about 0.008. This stays in agreement with the result of the Starostin's calculation, who predicted 0.0078  \cite{Starostin2003}. On the other hand, Starostin's calculation predicts the width of the $D2$ piece to be equal 0.0004 which is clearly on the verge of the resolution of our simulation, since there the edge length of the polygonal curve representing the rope was equal 0.00026. This explains why the $D2$ piece seen in Fig. \ref{SimpleClaspCurvature} is not fully developed. Knowing how narrow the $D1$ and $D2$ pieces are, one may immediately see that to find them in a numerical simulation is not a trivial task. That the SP-FEM method managed to complete the task, gave us more confidence both in its in organization and precision. 

Having reproduced the subtle details of the SS solution, we have also understood, that some unexpected, extremely narrow details that we have previously seen in the simulated tight knots, may be not artifacts of the discretization, but essential parts of the ideal conformations. The case of the tight overhand knot considered in ref. \cite{PrzybylPieranskiElasticOpenTrefoil} needs to be re-analyzed from this point of view using a higher precision simulation.

\section{The pursuit for the ideal trefoil knot: a retrospection}\label{Pursuit}

Looking into the history of the ideal trefoil problem we see how slowly, step by step, using various numerical knot tightening procedures such as: simulated inflation (SI) \cite{Katritch1996}, shrink-on-no-overlaps (SONO) \cite{PieranskiSONO1996}, deterministic ropelength minimizing algorithm (DRMA) \cite{Rawdon2003}, simulated annealing \cite{Carlen2005}, constrained gradient descent (ridgerunner) \cite{Ashton2011}, conformations of shorter and shorter ropelength have been found. (The {\it ropelength} of a knot is the ratio of the length of the rope in which it is tied to the radius of the rope. In terms of the notation used in the present paper, putting $R=1$ turns the length $L$ into ropelength.) Since, as a rule, at larger $N$ shorter conformations were found, together with shorter knots new details of the anatomy of the looked for ideal trefoil knot were discovered and new conjectures concerning its shape were formulated. It is the aim of the present paper to describe a few new details of this kind revealed in simulations carried with the use of the appropriately adapted finite element method (SP-FEM). The details are essential - they change our view on the shape of the ideal trefoil conformation. Before we present them, let us briefly summarize what up to now is known about the ropelength of the ideal trefoil knot. 

The described above hypothetical process, in which the knot becomes tightened because the rope in which it has been tied shrinks longitudinally while keeping its diameter intact or, keeping its length intact increases its diameter was for the first time realized in practice within the virtual reality of computer simulations by Katritch et al. \cite{Katritch1996}. The minimum length of trefoil knot determined in the simulations was equal $32.8$, although the method with which the number had been determined was not described in the paper. The first precisely defined value of the ropelength of the tight trefoil was given by Rawdon \cite{Rawdon2003} who developed a clear method with which the value should be calculated. The value given by Rawdon was equal $32.80$. An extensive study of the dependence of the ropelength $L$ of trefoil knots on the number of vertices $N$ used for their discretization was described by Baranska et al. \cite{Baranska2004}. The aim of the study was to estimate the ropelength of the tight polygonal trefoil knots found by the SONO algorithm in the $N \rightarrow \infty$ limit. To achieve this goal, trefoil knots with $N=1008, 1128, 1272, 1416, 1584, 1776, 2016, 2256$ and $2544$ vertices were tightened with the use of the SONO algorithm. Analyzing the dependence $L(N)$ the authors arrived to the conclusion that at $N \rightarrow \infty$ the ropelength should approach asymptotically the value $32.74295$. (Results of the study presented in the next section will allow us to comment this prediction.) Let us emphasize that the above value is but an approximate prediction. As far as the exact numerically provable values of the ropelength upper bounds are concerned, the length of the shortest trefoil knot was equal $32.7434$. It was found for a  knot with $N=2544$ vertices \cite{Baranska2004}. A slightly shorter knot with $N=3552$ vertices was found with the use of the SONO algorithm a few years later \cite{Baranska2008}. Its ropelength was equal $32.7432$. The shortest trefoils found by the \textsc{ridgerunner} algorithm \cite{Ashton2011} and by the simulated annealing technique \cite{Carlen2005} are a bit longer: 32.7437 and 32.7444, respectively. The knot that we are presenting below, contains $N=200640$ vertices and is shorter than any from the previously found knots. Its ropelength equals $32.7429345$, i.e. even less than the previously predicted $32.74295$ value \cite{Baranska2004}. Results of the pursuit for the determination of the ropelength of the ideal trefoil knot are gathered in table \ref{RopelengthTable}.

\begin{table}[ht]
\begin{ruledtabular}
\begin{tabular}{cccccc}

 & Author & Method & N & L &\\
\hline
 & Katritch et al. 1996 \cite{Katritch1996} & SI & 160 & 32.8  &\\
 & Pieranski et al. 2001 \cite{Pieranski2001} & SONO & 327 & 32.76 &\\
 & Rawdon 2003 \cite{Rawdon2003} & DRMA & 160 & 32.90 &\\
 & Baranska et al. 2004 \cite{Baranska2004}& SONO & 2544 & 32.7434 &\\
 & Carlen et al. 2006 \cite{Carlen2005}& SA & 528 (arcs) & 32.7444 &\\
 & Baranska et al. 2008	\cite{Baranska2008}& SONO & 3552 & 32.7432 &\\
 & Ashton et al. 2011 \cite{Ashton2011}& CGD & 2400 & 32.7437 &\\
 & Przybyl et al. 2012 & SP-FEM & 200640 & 32.7429345 &\\

\end{tabular}
\end{ruledtabular}

\caption{Ropelength values obtained by various authors. SI - simulated inflation, SONO - Shrink On No Overlaps, DRMA - deterministic ropelength minimizing algorithm, SA - simulated annealing, CGD - constrained gradient descent, SP-FEM - finite element method. Knots analyzed in papers \cite{Baranska2004} and \cite{Baranska2008} were found by Przybyl. Ropelength values marked with a question marked were calculated with a method that was not clearly specified.}
\label{RopelengthTable}
\end{table}

It may seem that the differences between the consecutive ropelength results presented above are so small that one should not bother about them. This is not the case. Although so small, the differences are by no means negligible. On the contrary, they are essential: finding a conformation with a slightly lower value of the ropelength we are getting closer to the ideal conformation and its portrait has a much better resolution. The most essential details of the ideal trefoil anatomy become visible only when the resolution is high enough. 

The ropelength is just a single parameter characterizing the tight knot found in a numerical simulation and it is of course desirable to know it: the lower the ropelength value, the closer the found knot to the ideal one. However, as a single geometrical parameter the ropelength does not tell us anything about the shape of the knot. To get an insight into the shape, we must look at the plots of curvature and torsion versus the arclength parameter: $\kappa(l)$ and $\tau(l)$. That the plots should be very precise indeed, since they may contain essential details of a very small size, we have demonstrated above considering the simple clasp structure.

Curvature and torsion functions give an exact and complete description of their shape. (To make the description unambiguous, the we must allow the torsion function to contain the Dirac-delta-like components.) Looking at plots of curvature and torsion one is able to indicate regions of the knot in which the rope has some well defined shapes, e.g. arcs or helices. The problem that we are facing is that torsion plots of the most tight conformations found by various knots tightening programs display a high level of noise. It is so, because numerical calculation of torsion needs the third derivative of the position vector field. 

One of the first papers in which essential conjectures concerning the shape the ideal trefoil knot were formulated was published by Carlen et al. \cite{Carlen2005}. Knots analyzed by the Lausanne team were approximated by bi-arcs. The most tight trefoil knot they managed to find after "{\it ... several months of the simulated annealing computation}" was built from 265 bi-arcs, thus, from $N=528$ arcs of varying length. The reported ropelength was $L=32.744$. The authors study in detail the knot aiming at formulating some conjectures concerning its shape. They calculate the radii of all arcs obtaining a plot, see figure 3(a) \cite{Carlen2005}, which reflects the curvature of the knot. Analysis of the plot supported by analysis of the contact set lead the authors to state what follows: 
\\
\\
{\bf Conjecture 1 (Carlen et al. \cite{Carlen2005})} 
{\it We remark that local radii are not active in the contact set for $\mu^* = 8.1861 \cdot 10^{-6}$, but are remarkably close to being active. For example local curvature does form part of the contact set $\chi_{5\mu^*}$. Thus it is quite possible that on the true ideal shape local curvature does achieve thickness at six distinct points}. 
\\
\\
Looking at the figure, one can see, that indeed one of the biarc curvature radii seems to reach the its lowest allowable value of the rope radius but at the five other minima the plot of the radii values remains clearly above the limit.
\\
As far as the torsion is concerned, the authors of ref. \cite{Carlen2005} did not plot its values, instead, they plotted unsigned values of the angles between planes of the osculating circles of the successive arcs. See figure 4 \cite{Carlen2005}. Analysis of this plot lead the authors to formulate the following conclusion: 
\\
\\
{\bf Conjecture 2  (Carlen et al. \cite{Carlen2005})} 
{\it There are three regions with large angles that correspond to the regions with a high variation of the radii, that is to the parts of the curve "inside the knot." Note that the angle is given in radians, that is, the maximal value of around 1.2 between adjacent arcs corresponds to an angle of around 70 degrees. These extremely large values could lead to the speculation that the Frenet frame of the underlying ideal curve may be discontinuous, and that the associated ideal centerline curve may not be $C^2$ at nearby points}.
\\
\\
Conjectures formulated by Carlen et al. \cite{Carlen2005} can be confronted with results of an analysis performed by Baranska et al. \cite{Baranska2008}. The knot analyzed by the authors contained $N=3550$ vertices. It was found after several hours of a simulation performed with the use of the SONO algorithm. The knot was highly equilateral: the relative deviation of the length of its edges from the average length was smaller than $10^{-7}$. Its ropelength calculated by the method developed by Rawdon \cite{Rawdon2003} was $L=32.74317$, thus the knot was more tight than the knot analyzed by Carlen et al. As clearly seen in Fig. 15 in ref. \cite{Baranska2008}, where the curvature of the $N=3552$ knot in the region of one of the double peaks, the first conjecture formulated by Carlen et al. is too weak. The curvature achieves indeed its upper limit value, but this happens not at six distinct points, but on six finite intervals. 
The speculation formulated by Carlen et al. concerning torsion of the ideal knot does not find in the analysis performed by Baranska et al. neither confirmation nor contradiction  \cite{Baranska2008}. The torsion data proved to be also very noisy. This forced the authors to use a smoothing procedure. The smoothing allowed them to plot torsion and indicate in the plot three double maxima, but excluded the possibility of proving or disproving the speculations concerning the continuity or discontinuity of the orientation of the Frenet frame. 
Another effort to find a tight conformation of the trefoil knot was made by Ashton et al. with the use of the Constrained Gradient Descent Method  \cite{Ashton2011}. The most tight trefoil knot they managed to find contained $N=2400$ vertices. Its ropelength $L$ was equal $32.743663$ which is more than the ropelength of the $N=3552$ trefoil found by Przybyl \cite{Baranska2008}. The result that the curvature hits its upper limit on six intervals has been confirmed. This is clearly seen in figure 13 of ref. \cite{Ashton2011}, where curvature is plotted. The question of the torsion plot was not considered.

\section{The ideal trefoil as seen in its $N=200640$ vertices portrait} \label{IdealTrefoil}

The extremely tight and precise polygonal trefoil knot $K_p$ containing $N=200640$ that we are going to present and analyze has been found by one of us (SP) in a simulation performed with the use of the appropriately modified finite element method (SP-FEM). Numerical calculations simulating the tightening of the knot lasted (on a PC computer) a few months and were finished when the knot reached the state at which no further tightening was possible. A simple numerical analysis reveals that the $K_p$ knot is highly symmetrical. Its symmetry is $D_3$. The knot has a single threefold symmetry axis and three, perpendicular to it, twofold axes. Thus, the knot consists of 6 congruent parts. In view of this, analyzing the shape of the knot we shall limit the analysis to a representative $1/6$ part of it. See Fig. \ref{IdealTrefoil}. To make plots of curvature and torsion that will appear during this analysis compatible with the plots presented above for the simple clasp, we have chosen as the starting point of the arclength axis $l$ the point at which the central minimum within the double peaks of curvature is located. The vertex has index equal 1. See Fig. 
 \ref{IdealTrefoil}.   

\begin{figure}[!ht]
	\centering
		\includegraphics[width=150pt]{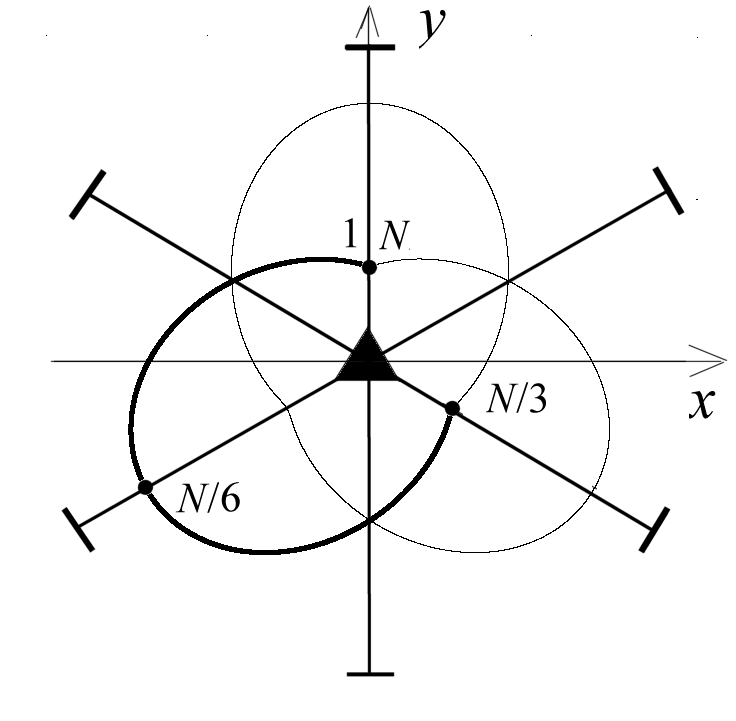}
	\caption{Vertex numbering and basic symmetry elements of the $K_p$ knot. Existence of the three- and two-fold symmetry axes allows one to limit analysis of the knot to $1/6$ part of it.}
	\label{IdealTrefoil}
\end{figure}

\subsection{Length}

Let the vertices of the polygonal knot $K_p$ be denoted by $\bm{v}_i$. The average distance between consecutive vertices, i.e. the average length of its edges $\bm{e}_i$
\begin{equation}
e_{av}=\frac{1}{N}\sum_{i=1}^N e_i = 1.63192456437\cdot 10^{-4},
\end{equation}
where
\begin{equation}
e_i=\left\|{\bm e}_i\right\|=\left \|{\bm v}_{i+1}-{\bm v}_i\right \|.
\end{equation}
Deviations of the segment lengths from the average value are not larger than $2\cdot10^{-15}$. In view of this, the $K_p$ knot can be treated as equilateral.

The length of the polygonal knot $K_p$ given with the accuracy stemming from the described above fluctuations of the length of its segments 
\begin{equation}
L_p=\sum_{i=1}^N e_i=32.742934460(1).
\end{equation}

Following Rawdon \cite{Rawdon2003}, the polygonal knot $K_p$ is replaced by the inscribed knot $K_c$. Since, as indicated above, the $K_p$ knot can be considered as equilateral, the construction of the Rawdon inscribed knot becomes particularly simple: consecutive arcs $a_i$ and $a_{i+1}$ inscribed at vertices $\bm{v}_i$ and $\bm{v}_{i+1}$ of the knot meet practically without any gap in the middle of the edge $\bm{e}_i=\bm{v}_{i+1}-\bm{v}_i$. 

Let $\Delta l_i,\; i=1,2,...,N$ be the lengths of the circular arcs, the union of which is $K_c$. The ropelength of the inscribed knot calculated with the Rawdon method gives:
\begin{equation}
L_c=\frac{\sum_{i=1}^N \Delta l_i}{R_c}=32.742934547(1),
\end{equation}
where
\begin{equation}
R_c=\sqrt{1-\frac{e_{av}^2}{4}}=0.99999999667(1).
\end{equation}
The given above $L_c$ value can be seen as a new, numerically provable, upper bound for the ropelength of the ideal trefoil. The previously known bound, equal $32.743386$, was given in ref. \cite{Baranska2004}. 

Calculations presented in ref. \cite{Baranska2004} show that the polygonal length underestimates the true ropelength of torus knots while the inscribed arcs length slightly overestimates it. The calculations indicate also that the true ropelength should be approximated well by an appropriately weighted average:
\begin{equation}
L_{a}=\frac{4}{5} L_p+\frac{1}{5} L_c
\end{equation}
Using this formula for the $K_p$ and $K_c$ knots we find:
\begin{equation}
L_{a}=32.742934477(1).
\end{equation}
The number can be seen as a new, more precise prediction of the ropelength of the ideal trefoil; the previous one, predicted in ref. \cite{Baranska2004} on the basis of knots tied with the SONO algorithm, was equal $32.74295$.  

\subsection{Curvature and torsion}

The ropelength of knot $K_c$, the shape of which we are going to analyze, is smaller than any from the previously found values. Thus, it is reasonable to assume that its structural details will be closer to the structural details of the ideal trefoil knot ${\cal K}_{id}$. 

As said above, at its full, original resolution, the curvature and torsion plots of discrete knots are not reliable since they contain small scale details that must be seen as artifacts of the discretization. Their origin has been explained in ref. \cite{Baranska2008}. The amplitude of the artifact details is reduced when, instead of the full set of vertices, we take into consideration $m$ times smaller set obtained by averaging coordinates of $m$ consecutive vertices. The knot obtained in such a manner will be denoted by $K_p^{(m)}$. The higher $m$, the more efficient the reduction of artifacts. On the other hand, $m$ cannot be too high since this would smear out all subtle details of the knot shape. As a rule we shall be using $m=10$. At this value of $m$ the number of vertices becomes reduced to $N^{(10)}=20064$ which is still about ten times more than the number of vertices in any of knots analyzed in the past. Averaging positions of $m=10$ consecutive vertices, makes the distance between them approximately $10$ times longer. Obviously, such a procedure introduces the dispersion of the edge length. However, since the original length of edges in the $K_p$ knot is extremely small, their dispersion in the $K_p^{(10)}$ knot is negligible. (The relative difference between the longest and shortest knot is of order $10^{-7}$.) Consequently, the Rawdon construction of the inscribed knot can be performed without any problems. Polygonal knot with the reduced number of vertices will be denoted $K_p^{(m)}$. The knot inscribed into it will be denoted by $K_c^{(m)}$. 

To grasp the shape of the whole knot one needs to analyze only one sixth part of it, i.e. for $l \in [0,L/6)$. In view of the very small difference between the lengths of the $K_c$ and $K_c^{(m)}$ knots, describing plots of curvature and torsion we are using the knot length parameter denoted simply as $L$.

\begin{figure}[!ht]
	\centering		\includegraphics[width=250pt]{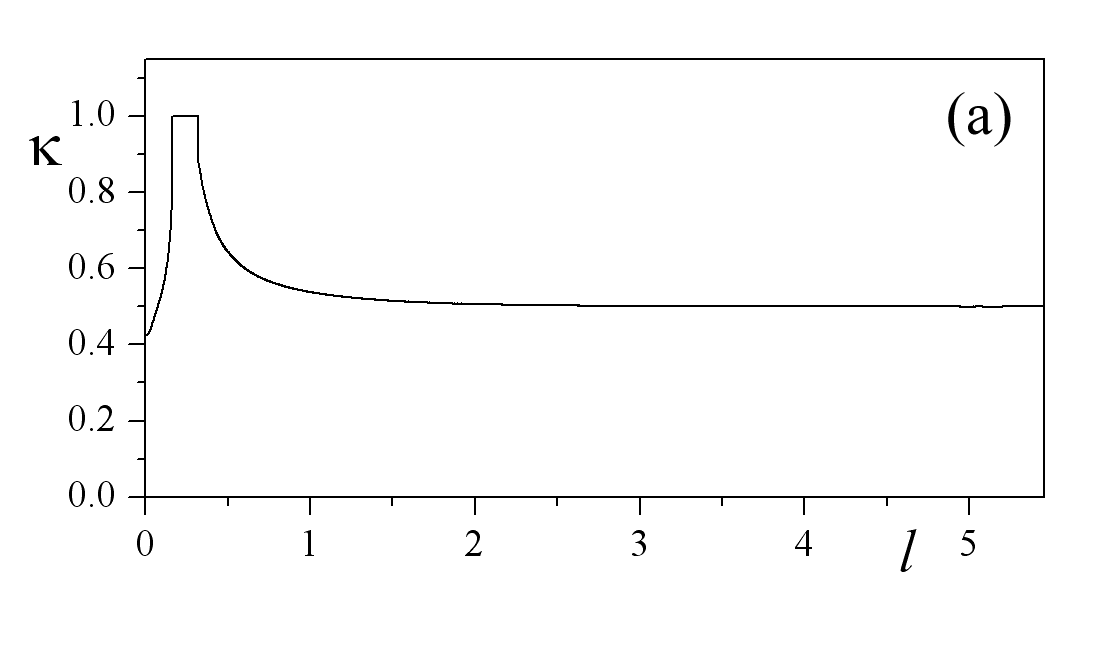}\\
	\centering		\includegraphics[width=250pt]{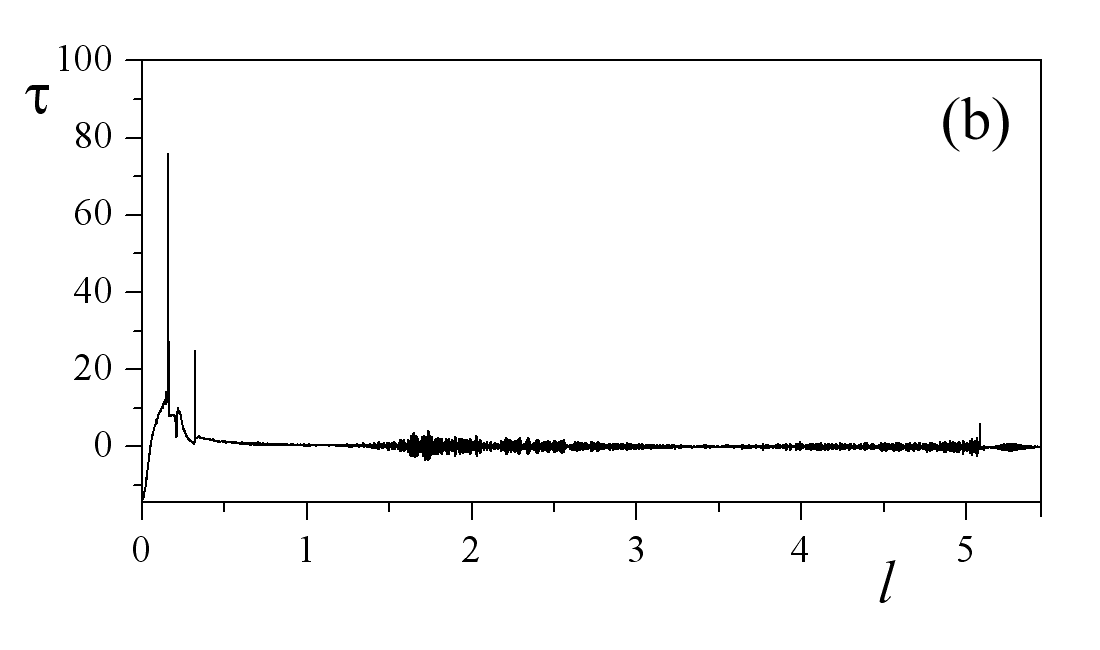}
	\caption{Curvature and torsion of the $K_c^{(10)}$ knot within the $[0,L/6)$ interval.}
	\label{CurvatureAll}
\end{figure}
	
\begin{figure}[!ht]
	\centering		\includegraphics[width=250pt]{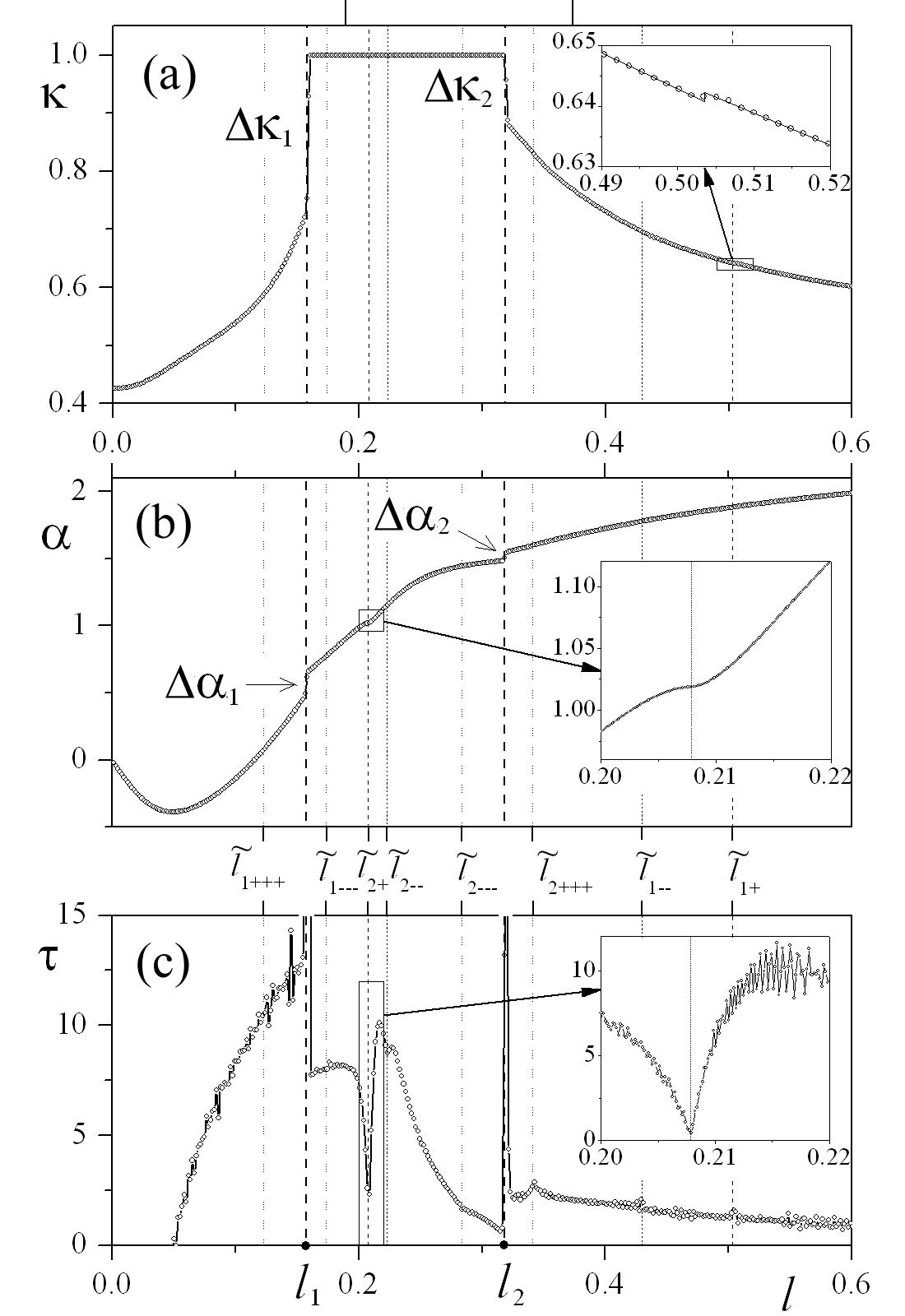}
	\caption{Curvature (a), accumulated torsion (b) and torsion (c) of the $K_c^{(10)}$ knot in the initial part of the $[0,L/6)$ interval. The curvature plot is clearly discontinuous at $l_1$ and $l_2$. The torsion plot displays there sharp and high peaks (their upper parts are not visible). As we guess, the peaks turn within the ideal trefoil knot into Dirac deltas. Note that values of torsion at both sides of the peaks are different.}
	\label{InitialPart}
\end{figure}

Figure \ref{CurvatureAll} presents the curvature plot of the  $K_c^{(10)}$ knot within the representative $1/6$ part of it. As clearly seen in the initial part of the plot, curvature reaches here its upper allowable value $\kappa=1$. This happens on a well defined, finite interval. The middle and the end parts of the curvature plot seem to be smooth. In what follows we shall demonstrate that it is not true.
The maximum curvature interval is separated from the rest of the plot by short pieces of very steep slopes. The steepness of the slopes becomes better visible when we look at the picture of the curvature plot enlarged in the interesting region. See Fig.\ref{InitialPart}(a). The shape of the curvature plot presented in the figure convinces us that we are dealing here with what in the ideal trefoil will appear as true discontinuities. In what follows the discontinuities will be referred to as \textit{primary}. We feel allowed to formulate the following:
\\
\\
{\bf Conjecture 1 (present authors)}
{\it Curvature of the ideal trefoil knot is not continuous. It reaches the maximum allowable value $\kappa=1$ on six finite intervals. The plateaus of maximum curvature are separated at both sides by discontinuities.}
\\
\\

The values of the arclength parameter at which the primary discontinuities of curvature appear are $l_1=0.158\pm 0.001$ and $l_2=0.319\pm 0.001$, thus the arclength width of the $\kappa=1$ interval equals $\Delta l_{1,2}=l_2-l_1=0.161 \pm 0.002$, which is more than $0.122$ found by Ashton et al. from the analysis of their $N=2400$ knot  \cite{Ashton2011}. The heights of the curvature jumps at both ends of each of the curvature peaks are: $\Delta \kappa_1 = 0.264 \pm 0.005$ and $\Delta \kappa_2 = 0.115 \pm 0.005$. Values of $\Delta \kappa_1$ and $\Delta \kappa_2$ cannot be confronted with results of other works, since the conjecture that curvature of the ideal trefoil knot is discontinuous was not formulated before.

Needing third derivatives of the position vector field, torsion of the $K_c^{(10)}$ knot is more difficult to determine. Figure \ref{CurvatureAll}(b) presents the torsion of the $K_c^{(10)}$ knot within the representative $1/6$ part of it. As seen in the figure, the level of noise is much higher than in the curvature plots. Nevertheless, in the interesting region of the curvature peak, where curvature displays discontinuities, one can easily notice two very sharp peaks. 
A study of the behavior of the torsion peaks at varying $m$ reveals that as $m$ decreases, the peaks become sharper and their height increases. Our conclusion is that, what the peaks represent are not conventional maxima of a smooth torsion function, but Dirac delta components of the function. The following hypothesis concerning torsion of the ideal trefoil seems to be thus well justified:
\\
\\ 
{\bf Conjecture 2 (present authors)}
{\it Torsion of the ideal trefoil is not a conventional function: in points, where curvature displays primary discontinuities, torsion displays Dirac delta components.}   
\\
\\
The weights of the Dirac delta components of torsion can be estimated by the inspection of the accumulated torsion $\alpha(l)$ plot. See Fig.\ref{InitialPart}(b). As seen in the figure, accumulated torsion is discontinuous at $l_1$ and $l_2$. Heights of the discontinuities are  $\Delta\alpha_1=0.159 \pm 0.005$ and $\Delta\alpha_2=0.062 \pm 0.005$. The values stay in a quantitative (but not qualitative) conflict with the observation made in ref. \cite{Carlen2005}, since it nowhere happens in the ideal trefoil knot {\it ...that the angle ... between adjacent arcs corresponds to an angle of around 70 degrees.} (70 degrees amounts to about 1.2 radians.) The weights of the Dirac delta components of torsion, i.e. the angles between the adjacent arcs at the places, at which as Carlen et al. conjectured the Frenet frame becomes discontinuous, are an order of magnitude smaller than $1.2$ radians.
A further detailed inspection of the torsion plot, see Fig.\ref{InitialPart}(c), reveals that values of torsion at both sides of the Dirac delta peaks are different. Thus, our conclusion is that at $l_1$ and $l_2$ torsion is also discontinuous.\\ 
There is an additional, unexpected feature of the torsion function. As seen in figure \ref{InitialPart}(c), inside the $(l_1,l_2)$ interval it displays a sharp, cusp shaped minimum located at point marked as $\tilde{l}_{2+}$. As we guess, the torsion function of the ideal trefoil knot is continuous but not smooth at the point. What is the origin of this singularity? To answer the question we must take into account contacts between points of the $K_p$ knot. 

\subsection{Contact set and contact functions}
\label{Contacts} 

\begin{figure}[!ht]
	\centering
		\includegraphics[width=200pt]{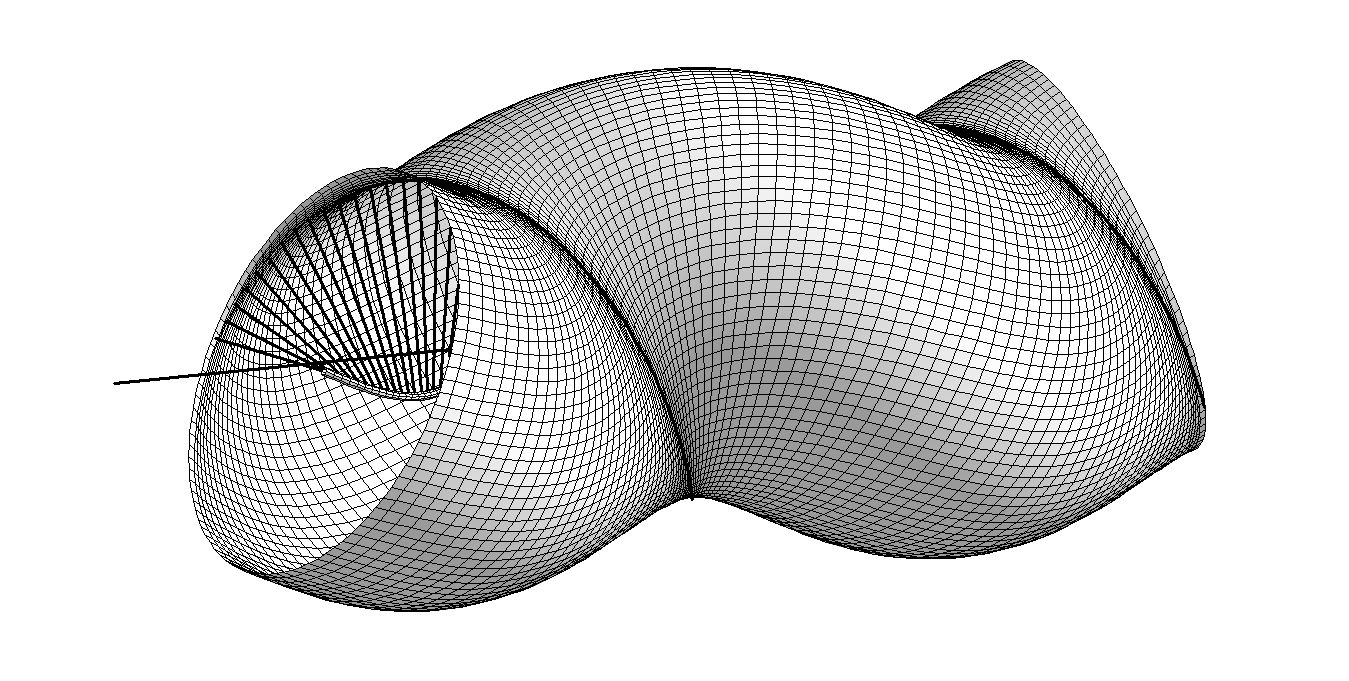}
	\caption{Curvature limited helix \cite{Przybyl2001}. The perfect rope of radius $R=1$ has been formed into a helix. Radius of the helix $r_H=0.2$, while its pitch $P_H=2.51327$. At such  parameter values the curvature of the helix is equal 1. Points of contacts of type II draw a helical line on the surface of the rope. The contact points coincide with centers of the osculating circles. Some radii that join the centers of the osculating circles with the points at which they are tangent to the helix are also shown.}
	\label{Helix}
\end{figure}

Following the notation used by Gerlach \cite{Gerlach2010}, let us denote by $\bm{\gamma}(l)$, $l \in [0,L_{id})$, the vector function describing the spatial positions of points along the looked for ideal trefoil knot ${\cal K}_{id}$. $L_{id}$ is here the length of the knot. 
Each point $\bm{\gamma}(l)$ of the ideal trefoil knot ${\cal K}_{id}$ can be seen as the center of a single, disk-shaped section $D(l)$ of the perfect rope in which the knot has been tightened. The section disks cannot overlap, but they may get into contacts. There are two kinds of the contacts:
\begin{enumerate}
\item  contacts between disks from two arclength distant parts of the rope,
\item  contacts between neighboring disks.
\end{enumerate}
The set of contact points of the first kind will be denoted by $CS_I$. The set of contacts of the second type will be denoted by $CS_{II}$. 
Pieces of the rope, within which contacts of the second type are present, are called kinks \cite{Cantarella2011}. Curvature reaches here its highest possible value $\kappa=1$. There are two kinds of kinks. In the first kind, present within the simple clasp structure discussed in Section \ref{Clasp}, section disks stay in contact at a single point. The axis of the rope remains here planar and can be seen as a circular arc of unit radius. In the second kind, the contact points between infinitesimally close disks are not gathered in the same place, but in a continuous manner are distributed along a curve. The simplest object within which such a curve of contacts of the second kind can be found is the perfect rope shaped into a curvature limited helix \cite{Przybyl2001}. See Fig. \ref{Helix}.
   
\begin{figure}[!ht]
	\centering
		\includegraphics[width=200pt]{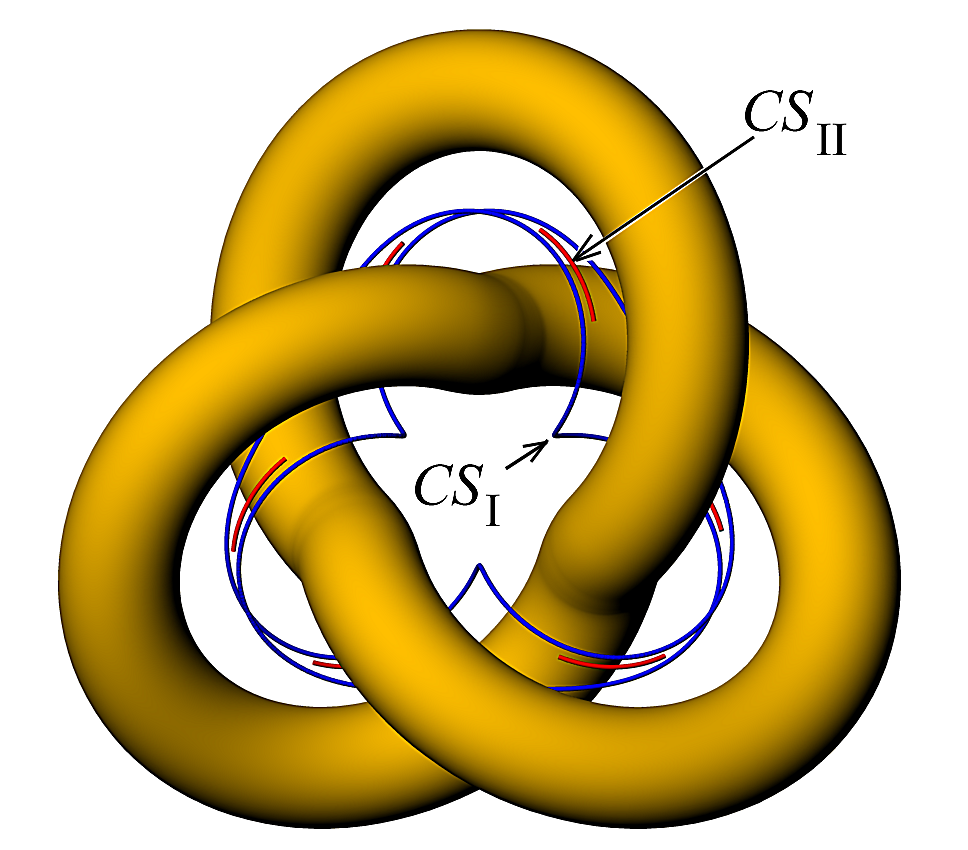}
	\caption{Conjectured location of the contact sets $CS_{I}$ and $CS_{II}$ within the ideal trefoil knot as determined via the analysis of the $K_p$ knot. Contact set $CS_{I}$ is connected and knotted. Contact set $CS_{II}$ consists of 6 congruent pieces. The arrow indicates one of them.}
	\label{ContactSets}
\end{figure}

Numerical simulations suggest, that each point of the ideal trefoil knot stays in contacts of the first kind with exactly two other points of the knot. Simulations described in \cite{Ashton2011}, \cite{Pieranski2001}, \cite{Gerlach2010}, \cite{Smutny}, \cite{Carlen2010} indicate the contact set $CS_I$ of the ideal trefoil knot is shaped into a continuous, self-avoiding, knotted curve. The contact set $CS_{II}$ has not been considered so far. The location of both contact sets within the knot structure is shown in Fig. \ref{ContactSets}. 

Points belonging to those parts of the knot, where $\kappa=1$, should be seen as having not two, but three contacts - the third one being of the second kind. This is essential since when considering the equilibrium of forces within the ideal trefoil knot (one may imagine that the perfect rope in which it has been tied is subject to tension), the presence of contacts of the second type must be also taken into account. From the physical point of view, the SP-FEM algorithm simulates a knotted, closed rope subject to a longitudinal tension. Acting along the rope the tension creates at its curved parts the transverse force that pushes vertices from different parts of the knot towards each other. The SP-FEM algorithm produces additional vertex-vertex forces aimed to prevent the distance between the vertices to become smaller than 2. On the other hand, the tension present within the simulated rope tries in some places to bend it more, than it is allowed. In such places, the SP-FEM algorithm produces additional momenta of forces that are preventing it. As a result, the inscribed knot $K_c$ is free from overlaps and its curvature nowhere exceeds 1.
Following notation used by Gerlach \cite{Gerlach2010}, we introduce two functions that allow one to indicate those points of the knot that get in contacts of the first type. Gerlach denotes the functions as $\sigma(l)$ and $\tau(l)$. Using the functions the pair of the knot points that get in contacts of the first type with a chosen point $\bm{\gamma}(l)$ of the knot can be specified as $\bm{\gamma}(\sigma(l))$ and $\bm{\gamma}(\tau(l))$. Since in accordance with a long tradition $\tau$ is used in the present paper as a symbol of torsion, we shall in what follows denote the Gerlach contact functions as $\sigma_-$ and $\sigma_+$. Using the notation, the points at which the contacts of the first kind take place are:
\begin{eqnarray}
\bm{c}_-(l)=\frac{\bm{\gamma}(l)+\bm{\gamma}(\sigma_-(l))}{2},\\
\bm{c}_+(l)=\frac{\bm{\gamma}(l)+\bm{\gamma}(\sigma_+(l))}{2}.
\end{eqnarray}
Obviously, to determine and parametrize the complete curve of the contact points $CS_I$ we need only one of the contact function, e.g.:
\begin{eqnarray}
CS_I(l)=\frac{\bm{\gamma}(l)+\bm{\gamma}(\sigma_-(l))}{2},\;l \in [0,L_{id}).
\end{eqnarray}

Let us return to the analysis of the polygonal $K_p$ knot found by the SP-FEM simulation and consider the problem of the contact functions. Approximate discrete images of the functions have been presented in references \cite{Ashton2011}, \cite{Pieranski2001}, \cite{Gerlach2010} and \cite{Smutny}. As mentioned above, the SP-FEM algorithm analyzes the vertex-vertex distances (the vertices belonging to different parts of the simulated rope) and whenever they become smaller than 2, it introduces a pair of opposite forces aiming to remove the violations. From the physical point of view, two vertices between which the forces are acting can be seen as connected with a strut subject to a compression. When in the course of the SP-FEM simulation the forces acting on a strut change sign, the strut is removed. Thus, at the end of the simulation, the simulation code provides us with the set $B$ of the pairs of vertices between which the compressed, distance keeping struts are present. Vertices connected by a strut can be seen as staying in a contact of the first kind. The contact point is located in the middle of the strut. Numerical analysis shows that the absolute deviation of the strut lengths from their desired value 2 is not larger than $8 \cdot 10^{-15}$. 

\begin{figure}[!ht]
	\centering
		\includegraphics[width=150pt]{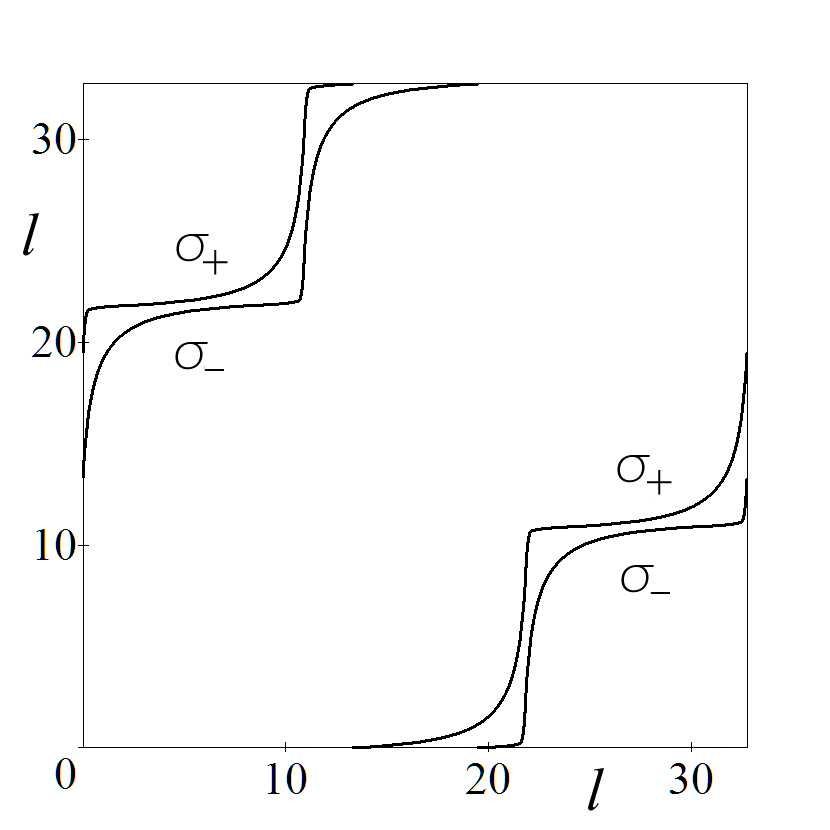}
	\caption{Shapes of the contact functions $\sigma_-$ and $\sigma_+$ as seen via plots of the end points of the contact struts.}
	\label{ContactFunctions}
\end{figure}

Let $\left\{i^B_k,j^B_k\right\}$, $k=1,2,...,M$, be the pairs of indices defining the struts in the final $K_p$ knot. Ordered pairs $(l_{i^B_k},l_{j^B_k})$ and $(l_{j^B_k},l_{i^B_k})$ define within the $[0,L_p]\times[0,L_p]$ square points that can be seen as pixels of a high but finite resolution, discrete image of the continuous contact functions $\sigma_-(l)$ and $\sigma_+(l)$. See Fig. \ref{ContactFunctions}. An appropriate processing of the data allows us to find values of the strictly monotone $\sigma_-(l)$ and $\sigma_+(l)$ functions. Analyzing the functions we are able to draw some new conclusions concerning their properties. 
It is essential to notice that in view of the conjectured $D_3$ symmetry of the ideal trefoil knot to know the shape of both contact functions in the whole $[0,L_{id}$ range it is sufficient to know the shape of one of them in the initial $[0,L_{id}/3)$ interval or the shapes of both functions in the $[0,L_{id}/6)$ interval. 
   
\begin{figure}[!ht]
	\centering		\includegraphics[width=250pt]{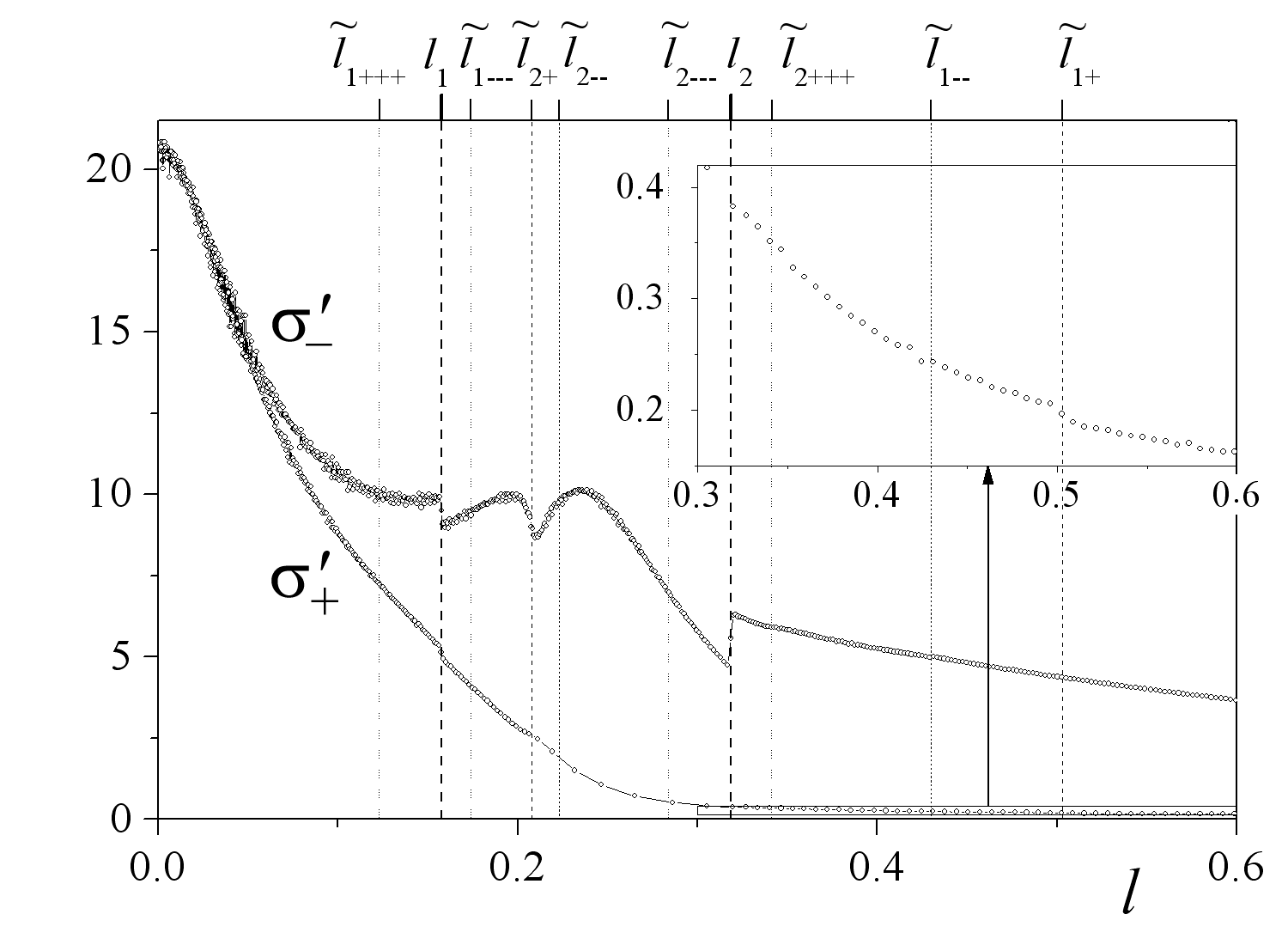}
	\caption{Shape of the derivatives of the contact functions $\sigma_-$ and $\sigma_+$ in the initial part of the $[0,L_c/6)$ interval. }
	\label{DerivativeOfContactFunctionsInitialPart}
\end{figure}

\begin{figure}[!ht]
	\centering		\includegraphics[width=250pt]{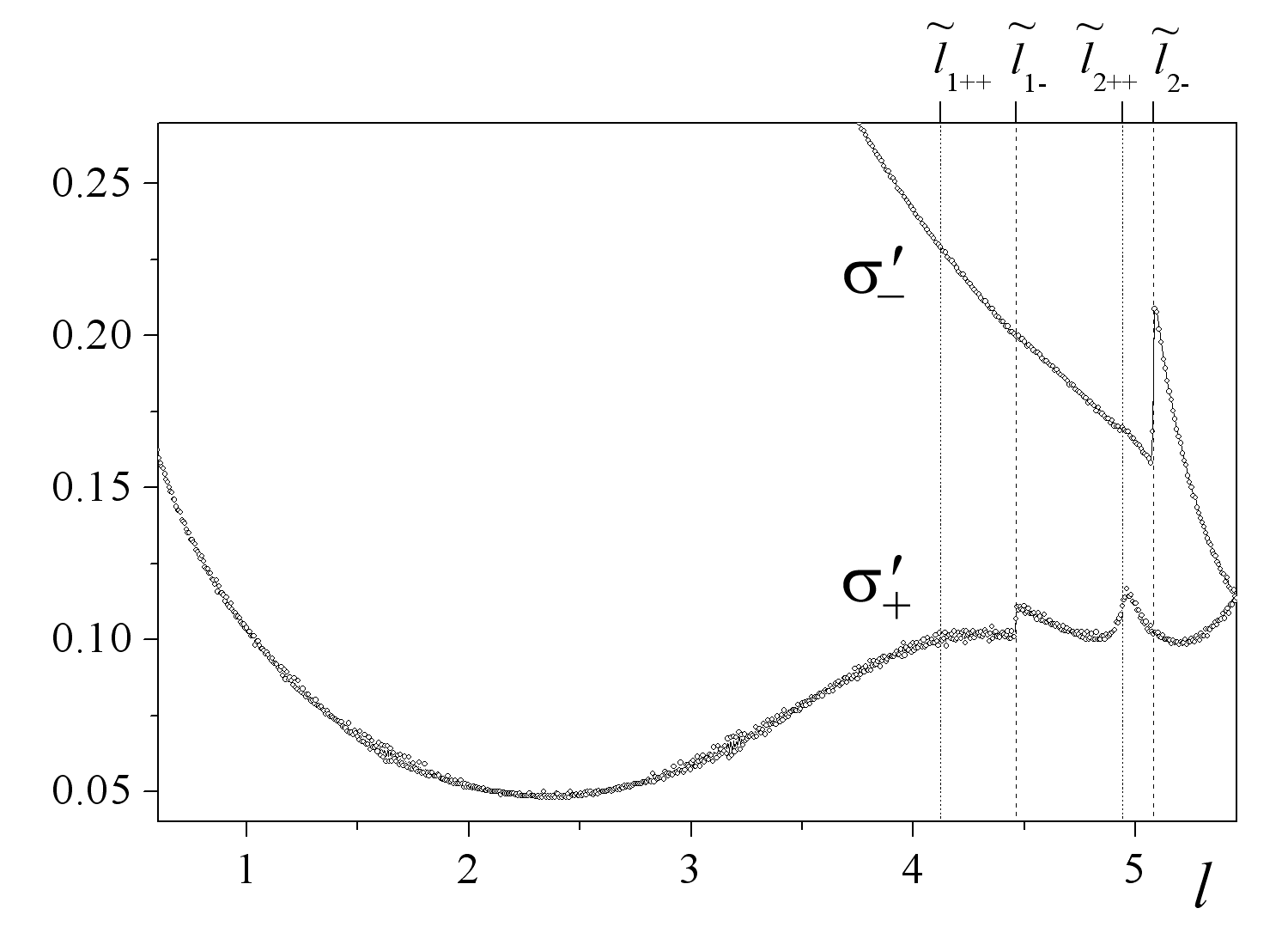}
	\caption{Shape of the derivatives of the contact functions $\sigma_-$ and $\sigma_+$ in the end part of the $[0,L_c/6)$ interval. }
	\label{DerivativeOfContactFunctionsEndPart}
\end{figure}

Needing it for the proof that the contact curve is knotted Gerlach formulates a hypothesis (5.14 in ref. \cite{Gerlach2010}), that the contact functions are smooth. Inspection of the plots presented in references \cite{Ashton2011}, \cite{Pieranski2001}, \cite{Gerlach2010}, \cite{Smutny} and \cite{Carlen2010} seems to confirm it. At the first sight, see Fig.\ref{ContactFunctions}, results of the present study stay also in agreement with the hypothesis. However, inspecting the derivatives of the functions we clearly see that it is not: see Fig. \ref{DerivativeOfContactFunctionsInitialPart} and Fig. \ref{DerivativeOfContactFunctionsEndPart}. 
The most distinct  discontinuities of the derivative of the $\sigma_-$ function are located at $l_1$ and $l_2$ where curvature displays its primary discontinuities. It is clear, that the function is not smooth also at other points. Analysis of the positions and values of the discontinuities and the relations between them needs a separate study.  

\begin{figure}[!ht]
	\centering
		\includegraphics[width=200pt]{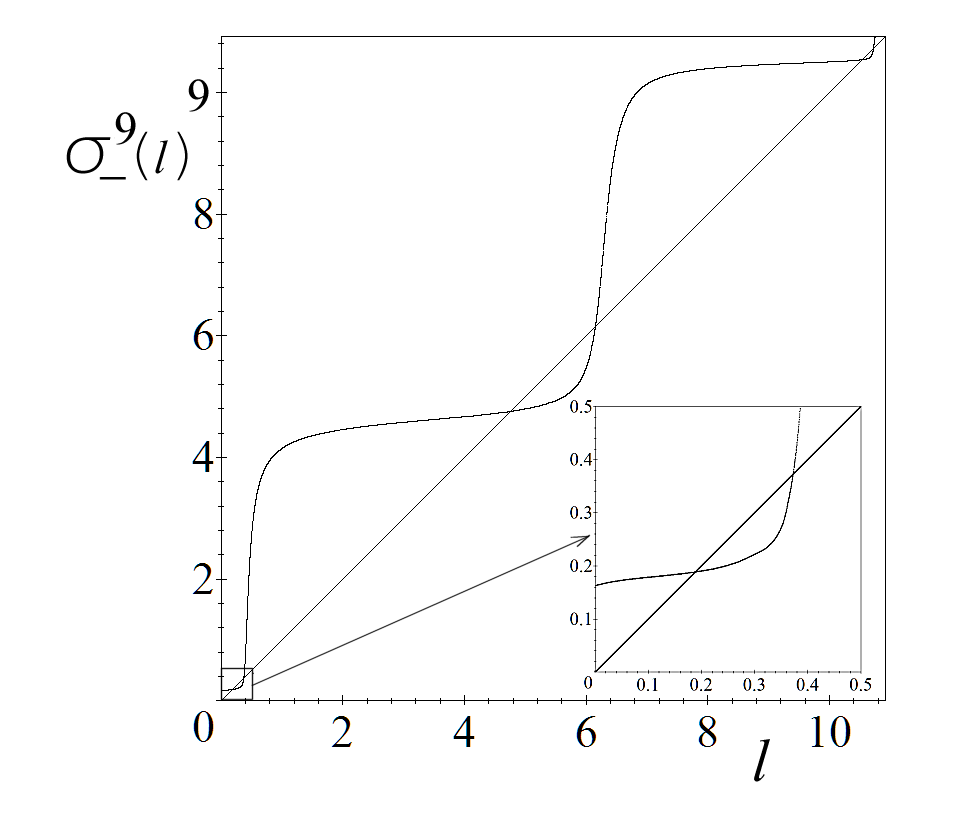}
	\caption{The shape of the 9-th functional power $\sigma_-^9$ of the contact function $\sigma_-$ determined from the SP-FEM struts data. To make the intersection points of the $\sigma_-^9$ plot with the diagonal we plotted but $1/3$ part of the function. There are here 6 intersection points, thus, in the whole knot there are 18 of them.}
\label{sigma9}
\end{figure}

\begin{figure}[!ht]
	\centering
		\includegraphics[width=200pt]{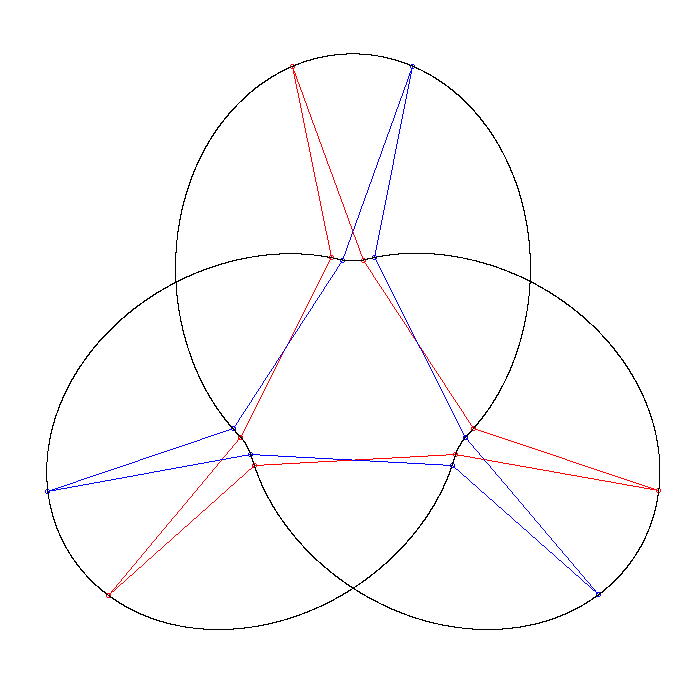}
	\caption{Conjectured position of two 9-cycles within the ideal trefoil knot.}
\label{Cycles9}
\end{figure}

Another of the hypotheses formulated both by Gerlach and Carlen \cite{Gerlach2010}, \cite{Carlen2010} is the existence within the ideal trefoil of the so-called 9-cycles. A plot of the $\sigma^9$ function obtained with the data stemming from the $K_p$ knot indicates that it is not tangent to the diagonal, but intersects it in 18 points. See Fig. \ref{sigma9}. The 18 points form two stable 9-cycles. The arclength positions of the points of the first and second cycle (listed in the order in which they are connected in a given cycle via consecutive actions of the contact function $\sigma_-$) are: $C^9_-=$[0.189, 15.674, 32.370, 11.103, 26.588, 10.541, 22.018, 4.759, 21.455] and $C^9_+=$[ 32.554, 11.288, 27.984, 10.725, 22.202, 6.155, 21.640, 0.373, 17.069]. We denoted the two cycles by $C^9_-$ and $C^9_+$ since although map $l\rightarrow \sigma^9_-(l)$ has 18 fixed points only 9 of them, gathered in $C^9_-$, are stable. The other 9 fixed points, gathered in $C^9_+$, prove to be stable from the point of view of mapping  $l\rightarrow \sigma^9_+(l)$.\\
Location of the 9-cycles within the spatial structure of the knot is shown in Fig. \ref{Cycles9}. As Carlen indicates \cite{Carlen2010}, assuming that symmetry of the ideal trefoil is $D_3$, the existence of a single 9-cycle allows one to construct the whole knot from only two pieces of it. Results of the present study leads to a modification of the statement. Since, as we have demonstrated, we deal not with one but with two 9-cycles, the construction of the whole knot needs not two but three pieces \cite{TwoCycles}. 

\section{Higher order singularities of curvature and torsion}

Assuming that the ideal knot has the $D_3$ symmetry and choosing properly the place at which the arclength $l$ variable has its origin, we come to the conclusion that what happens in the knot at a point $\bm{\gamma}(l)$, where $l \in [0,L_{id}/6)$, happens also at five other points $\bm{\gamma}(L_{id}/3-l)$, $\bm{\gamma}(L_{id}/3+l)$, $\bm{\gamma}(2L_{id}/3-l)$, $\bm{\gamma}(2L_{id}/3+l)$, and $\bm{\gamma}(L_{id}-l)$. Obviously, the inverse is also true: whatever happens at $\bm{\gamma}(l)$, where $l \in [L_{id}/6,L_i)$, happens also at a $\bm{\gamma}(\tilde{l})$, where $\tilde{l} \in [0,L_{id}/6)$. Formulae given above make it clear how to calculate the $\tilde{l}$ value:
\begin{equation}
\tilde{l}=\begin{cases}
 l & \text{for $ l\in [0/L_{id}/6)$ (I)} \\ 
 2L_{id}/6-l & \text{for $ l\in [L_{id}/6,2L_{id}/6)$ (II)} \\
 l-2L_{id}/6 & \text{for $ l\in [2L_{id}/6,3L_{id}/6)$ (III)} \\
 4L_{id}/6-l & \text{for $ l\in [3L_{id}/6,4L_{id}/6)$ (IV)} \\
 l-4L_{id}/6 & \text{for $ l\in [4L_{id}/6,5L_{id}/6)$ (V)} \\ 
 L_{id}-l & \text{for $ l\in [5L_{id}/6,L_{id})$ (VI)} \\
 \end{cases}   
\end{equation}
 
\begin{figure}[!ht]
	\centering		\includegraphics[width=250pt]{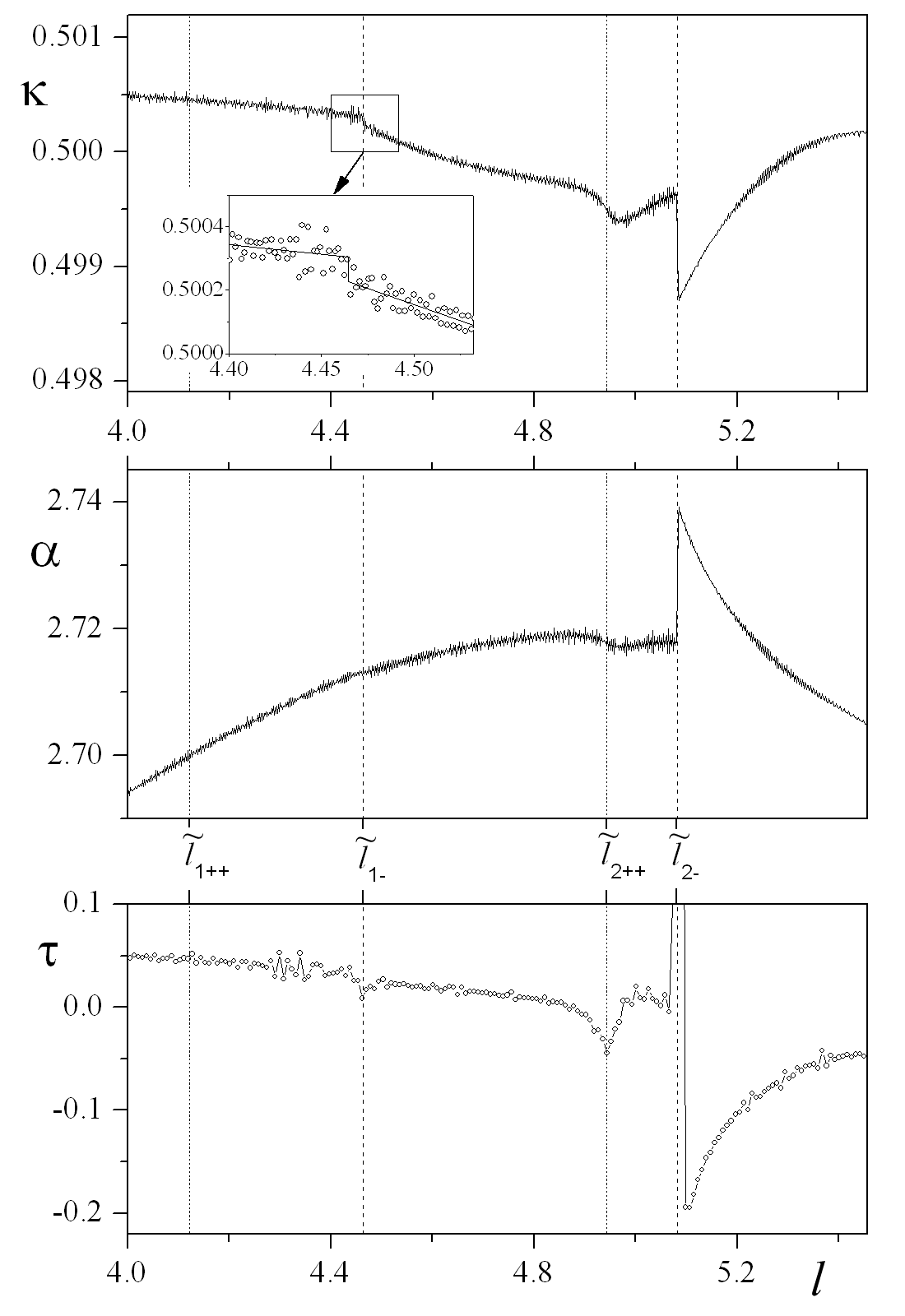}
	\caption{Enlarged plots of curvature (a), accumulated torsion (b) and torsion(c), in the end part of the $[0,L/6)$ interval. The secondary discontinuity of curvature visible at $\tilde{l}_{2-}$ is induced by the primary discontinuity visible in Fig.\ref{InitialPart}(a) at $l_2$. }
	\label{EndPart}
\end{figure}

Suppose we start from point $l_2=0.3185$ (located within the $[0,L_{id}/6)$ interval) at which the primary discontinuity of curvature is takes place. The $l_2$ point stays in contact with point $\l_{2-}=\sigma_-(2_1)=16.7467$. Since $l_{2-}\in [3L_{id}/6,4L_{id}/6)$, $\tilde{l}_{2-}=4L_{id}/6-l_{2-}=5.0819$. Thus, the influence of the primary discontinuity of curvature located at $l_2$ on the shape of curvature at the $\sigma_-(2_1)$ should visible in the representative $[0,L_{id}/6)$ interval at $\tilde{l}_{2-}$. Inspection of the curvature plot presented in Fig.\ref{CurvatureAll} is disappointing, since at this value of $l$ the curvature plot seems to be smooth. However, an appropriate magnification reveals the interesting details. As seen in the Fig.\ref{EndPart}, curvature is also discontinuous at the point. This discontinuity will be called \textit{secondary}. Its height equals  $0.00095 \pm 0.00008$. Further inspection of the figure indicates that the accumulated torsion, part (b) of the figure, is also discontinuous. Thus, torsion shown in part (c) should have a Dirac delta component. Indeed, at $\tilde{l}_{2-}$ torsion displays at a very high peak. The height of the peak depends strongly on $m$, thus it represents a Dirac delta component. The weight of the component (equal to the height of the accumulated torsion discontinuity) amounts to $0.0215 \pm 0.0016$.

Continuing the analysis we may ask what influence has the $l_2$ discontinuity on the shape of the knot at the point indicated by the second, i.e. $\sigma_+$, contact function. The place within the $[0,L/6)$ interval, where it should be seen is located at $\tilde{l}_{2+}=0.2079$. As seen in Fig.\ref{InitialPart}(a) the point falls into the $(l_1,l_2)$ interval, where curvature is limited by its upper allowable value $\kappa=1$. Thus, looking at the curvature plot, we do not see anything. Inspection of the accumulated torsion and torsion plots, (b) and (c), reveals that the contact leaves a visible trace: the accumulated torsion has a kind of an inflection point there. Torsion displays at the point a sharp, cusp shaped minimum. Our conjecture is that at the $\tilde{l}_{2+}$ point of the ideal trefoil knot its torsion is continuous but not smooth. It reaches there the zero value.\\
Similar analysis allows one to indicate the origin of other interesting landmarks visible in the curvature and torsion plots. For instance, the secondary discontinuity of curvature shown in the picture inserted in Fig.\ref{InitialPart}(a) is located at $\tilde{l}_{1+}=0.5029$ point. Its height amounts to $0.0016 \pm 0.0008$, thus it is very small. The secondary discontinuity shown in the picture inserted in Fig.\ref{EndPart}(a) is located at $\tilde{l}_{1-}=4.46361$  and its height is hardly visible: $0.00005 \pm 0.00005$. Ternary singularities located at points $\tilde{l}_{1--}$ and $\tilde{l}_{1++}$ are much weaker. It seems that at $\tilde{l}_{2--}$ and $\tilde{l}_{2++}$ there are no singularities, since $\tilde{l}_{2+}$ is located within the interval of constant curvature and the same happens with $\tilde{l}_{2--}$. We managed to detect some of them, but since their magnitudes are burdened with large errors, we do not list them.

\section{Conclusions}

Results of the presented above simulations allowed us to formulate firmly supported conjectures concerning the shape of the ideal conformation of the closed trefoil knot. Let us gather them into a single list:

1. Curvature $\kappa$ of the ideal trefoil is not continuous. The primary discontinuities appear at the ends of the intervals within which $\kappa=1$.  

2. Torsion $\tau$ of the ideal trefoil knot is not continuous. Localization of its singularities is identical with localization of the discontinuities of curvature. At points of the primary discontinuities of curvature, the torsion function has singular Dirac delta components. 

3. Contact functions of the ideal trefoil knot are not smooth. The points of their discontinuities are related to the discontinuities of curvature and torsion.

4. The ideal trefoil knot hosts two 9-cycles. 

5. Contacts between various parts of the ideal trefoil knot propagate each of its primary shape singularities along the knot, but the existence of the constant curvature intervals and the 9-cycles makes the propagation process to decay.

These new, listed above conjectures change essentially our view on the shape of the ideal trefoil knot, indicating that it is more complex than it has been assumed so far. Let us emphasize that acceptance of the existence in the ideal knot shape singularities was possible only because we were keeping in mind results of the Starostin and Sullivan study of the simple clasp. In a similar manner, the existence of the secondary shape singularities and their connections with the primary singularities became clear only due to the idea of the contact functions introduced by Carlen and Gerlach.
The curvature, torsion and contact funtions singularities are closely related. A general study of the connections between them still needs some more work.\\
     
\section{Acknowledgments}

John Sullivan and Evgueni Starostin drew our attention to the simple clasp problem and to its analytical solution. They patiently explained to us the importance of the subtle details of its structure. Jason Cantarella drew our attention to the problem of local minima. Eric Rawdon taught us how to analyze properly the polygonal knots produced by SONO and SP-FEM. PhD theses of Henryk Gerlach and Mathias Carlen helped us to understand the importance of the contact functions and indicated possibility of the existence of the 9-cycles. Rob Kusner gave us the moral support necessary to complete the work. We thank all of them. The work was supported by Poznan University of Technology Grant 06/62/DSPB/0214/2014.


\begin{thebibliography}{10}
\bibitem{TwoCycles} The old data analyzed in PhD theses by Gerlach and Carlen suggested the existence of a single 9-cycle, but their new data indicate that there are two such cycles.
\end{thebibliography}
\end{document}